\documentclass[
aps,
pra,
twocolumn,
groupedaddress,
longbibliography
]{revtex4-1}
\usepackage{graphicx}
\usepackage{amsmath}
\usepackage{amssymb}
\usepackage{siunitx}

\DeclareMathAlphabet{\bi}{OML}{cmm}{b}{it}

\newcommand{\rmi}{\mathrm{i}}

\newcommand{\imagi}{\rmi}

\newcommand{\diff}{\mathrm{d}}

\newcommand{\pabl}[2]{\frac{\partial #1}{\partial #2}}

\newcommand{\halb}{\frac{1}{2}}

\newcommand{\beq}{\begin{equation}}
\newcommand{\eeq}{\end{equation}}

\newcommand{\Wcmcm}{W/cm$^2$}

\begin{document}

%\preprint{}

\title{Robustness of topologically sensitive harmonic generation in laser-driven linear chains}


\author{Helena Dr{\"u}eke}
%\thanks{}
\author{Dieter Bauer}
\affiliation{Institute of Physics, University of Rostock, 18051 Rostock, Germany}


\date{\today}

\begin{abstract}
A many-order-of-magnitude difference in the harmonic yield from the two topological phases of finite, dimerizing linear chains in laser fields has recently been observed in all-electron time-dependent density functional simulations [D.\ Bauer, K.\ K.\ Hansen, Phys.\ Rev.\ Lett.\ {\bf 120}, 177401 (2018)].
In this work, we explore the robustness of the effect concerning the length of the chains, a continuous transition between the two topological phases, and disorder.
A high robustness of both the degeneracy of the edge states in the topologically non-trivial phase as well as of the pronounced destructive interference, causing a dip in the harmonic spectra, in the topologically trivial phase is observed. 
\end{abstract}

\pacs{}
%\keywords{}

\maketitle

% References should be done using the~\cite, \ref, and \label commands
\section{Introduction}
The exploration of the territory between strong-field short-pulse laser and condensed matter physics has only recently gained momentum.
 Electron dynamics~\cite{Schultze1348,  Lucchini916, Hassan2016, Sommer2016} and high harmonic generation~\cite{Ghimire2011, SchubertO.2014, Hohenleutner2015, VampaPhysRevLett.115.193603, Luu2015, LangerF.2017, TancPhysRevLett.118.087403, Vampa2018, Garg2018}
are studied in crystals as well as amorphous solids~\cite{You2017}.
Studies on noble gases allow for comparison of high harmonics produced by gases and their corresponding solids~\cite{Ndabashimiye2016}.
Furthermore, first investigations on two-dimensional solids, such as monolayer graphene~\cite{Higuchi2017, Baudisch2018, PhysRevLett.121.207401} or ferromagnetic monolayers~\cite{Zhang2018} are being performed.

High harmonic generation in gases can be explained by the three-step-model~\cite{corkum_plasma_1993,LewensteinPhysRevA.49.2117}.
In the first step, an electron tunnels into the continuum.
The electron is then accelerated by the oscillatory laser field in the second step and possibly driven back to its parent ion.
In the third step, the electron may recombine with the ion upon emission of  high-harmonic radiation. 
The three-step-model can be adapted for solids by taking the band structure into account. The atomic ground state is replaced by the valence band, the continuum by the conduction bands, and the recombination of the electron with its parent ion by the recombination of the electron moving in one of the conduction bands with the hole moving in the valence band \cite{VampaTutorial0953-4075-50-8-083001}. This so-called interband harmonic generation can well explain the cut-offs observed in ab-initio simulations (see, e.g., \cite{WuPhysRevA.94.063403,PhysRevA.96.053418}).
Intraband harmonics, on the other hand, are caused by electrons moving within bands.
Due to the varying curvature of the bands, this motion generates harmonics as well. These harmonics are expected to be dominant for harmonic energies below the band gap between the lowest (laser-dressed) conduction band and the valence band because the three-step mechanism does not generate such sub-band-gap harmonics.
One might expect that a fully occupied valence band does not produce intraband harmonics. However, in a simulation with non-interacting (Kohn-Sham) electrons, all electrons move and, individually, generate harmonics. It is only by destructive interference that the total harmonic yield largely cancels.  

One of the fascinating research directions in condensed matter physics, cold atoms, and photonics are topological phases  with interesting properties, and transitions between them~\cite{1367-2630-18-8-080201}.
From the strong-field laser perspective, there are only very few investigations on that subject to date~\cite{PhysRevB.96.075409,Luu2018,bauer_high-harmonic_2018,silva_all_2018,chacon_observing_2018}.
Many questions arise in that context, for instance:
How are topological phases encoded in typical strong-field observables such as photoelectron or harmonic spectra?
How can strong, short-pulse lasers be used to manipulate topological phases?
May topological edge states be useful to develop ultrafast, light-driven electronic devices or, vice versa,  electronically driven, novel light sources?
And from the theoretical perspective: Is the usual tight-binding modeling enough to  properly describe the nonlinear light-matter interaction? Which are the suitable and useful topological invariants?
Does the laser need to illuminate the edges to have topological edge effects in photoelectron or harmonic spectra?

\begin{figure}[htbp]
 \centering
 \includegraphics[width = 0.7\columnwidth, keepaspectratio]{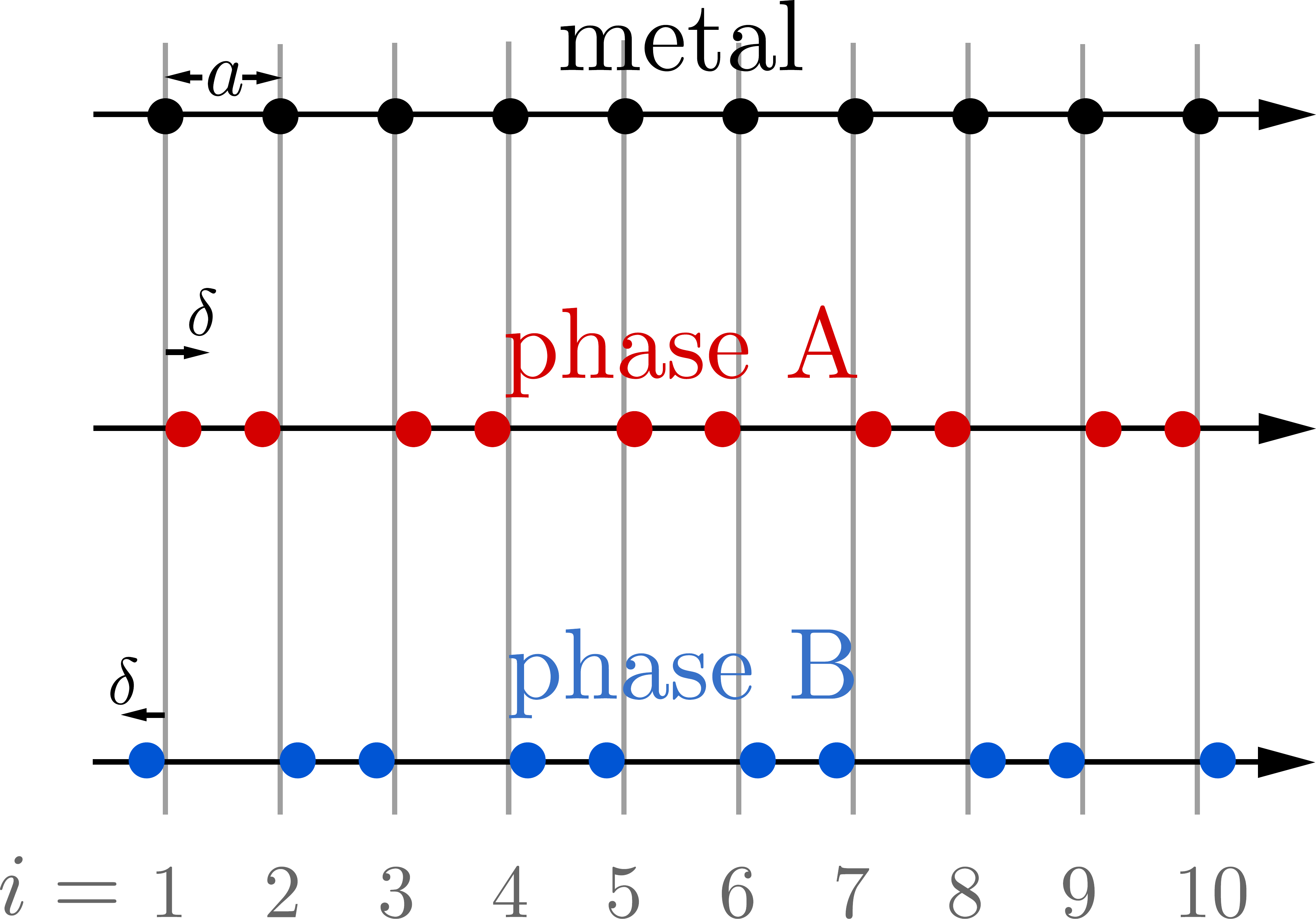}
 \caption{Sketch of the ion positions for a dimerizing chain of length $N=10$: equidistant separations $a$ (metal), ions shifted alternatingly to the right and the left ($\delta > 0$, phase~A), ions shifted alternatingly to the left and to the right ($\delta < 0$, phase~B).}
 \label{fig:positions_phases}
\end{figure}

In Ref.~\cite{bauer_high-harmonic_2018}, a many-order-of-magnitude different harmonic yield is observed in the sub-band gap regime for the two topological phases of dimerizing linear chains. Figure~\ref{fig:positions_phases} illustrates the setup. A linear chain of equidistantly separated atoms with one electron per atom has a half-filled valence band and thus is a metal. By shifting the atoms by $\delta$ alternatingly to the left and to the right (or vice versa), pairs of atoms with a shorter distance $a-|\delta|$ between each other are formed. In the band structure, a gap opens, the total energy decreases (Peierls instability),and the system becomes insulating. However, in so-called phase A all atoms are paired whereas in phase B the two edge atoms remain unpaired. This difference may look trivial at first but has profound consequences, which are most clearly elucidated using the Su-Schrieffer-Heeger (SSH) model~\cite{SSHPhysRevLett.42.1698,topinsshortcourse}. The SSH model is a tight-binding version of the situation depicted in Fig.~\ref{fig:positions_phases}: electrons hop with higher probability between the paired sites and with lower probability between the more distant sites. The SSH model for the bulk is simple enough to allow for the definition of a winding number as a topological invariant. It turns out that phase A is topologically trivial while phase B has a non-vanishing winding number \cite{topinsshortcourse}. The simulations in Ref.~\cite{bauer_high-harmonic_2018} are based on time-dependent density functional theory (TDDFT)~\cite{runge_density-functional_1984,UllrichBook}, which goes far beyond a tight-binding, independent-electron treatment. Nevertheless, the main topological feature of the SSH model---the presence of degenerate topological edge states---is ``inherited'' by the less idealized TDDFT description. The difference in the harmonic yield (also seen in Fig.~\ref{fig:HHG_fewer_ions} below) is attributed to pronounced destructive interference of all the electrons' intraband harmonic emission from the valence band of the topologically trivial phase~A while the destructive interference is spoiled by the presence of half-occupied edge states in the topologically non-trivial phase~B (with edge ions).

From a practical point of view, robustness is the most appealing asset of topological matter because of potential applications, for instance in quantum computation (topological qubits~\cite{Stern1179}) or high-temperature superconductivity~\cite{Fatemieaar4642}.
In the present paper, we investigate the robustness of the topological effect in harmonic generation observed in~\cite{bauer_high-harmonic_2018} with respect to chain size, continuous phase transition, and disorder.

The paper is organized as follows:
We review the TDDFT model in Section~\ref{sec:model}.
The robustness of the degeneracy of the edge states and the harmonic spectra is investigated in Section~\ref{sec:robust} with regard to size dependence (subsection~\ref{sec:size}), a continuous phase transition (subsection~\ref{sec:trans}), and disorder in the ion positions and ion potentials (subsections~ \ref{sec:disorderpos} and \ref{sec:disorderpot}, respectively).
We summarize in Section~\ref{sec:summ}.

Throughout this paper, atomic units $\hbar = |e| = m_\mathrm{e} = 4 \pi \epsilon_0 = 1$ are used unless stated otherwise.

\section{Density-functional model for linear chains}\label{sec:model}
The model system used in this work is the same as in~\cite{bauer_high-harmonic_2018}, a linear chain of $N$ singly charged ions.
Starting from an equidistant spacing, the ions are alternatingly shifted by $\delta$ to the left and to the right (see Fig.~\ref{fig:positions_phases}), leading to ion positions
\beq x_i = \left(i - \frac{N + 1}{2}\right) a - (-1)^i\delta, \qquad i=1,2,\ldots,N \eeq 
with the lattice constant $a$ and shift $\delta$.
The interaction of an electron (at position $x$) with the ions is described by the sum of the soft-core Coulomb potentials
\beq v_\mathrm{ions}(x) = \sum_{i=1}^{N} v_i(x) = - \sum_{i=1}^{N} \frac{1}{\sqrt{(x-x_i)^2 + \varepsilon_i}}.\eeq 
We choose the same smoothing parameter for all ions, $\varepsilon_i = 1$, except for subsection \ref{sec:disorderpot}, where topological robustness with respect to random fluctuations of the $\varepsilon_i$ is investigated.

We use time-dependent density functional theory~\cite{runge_density-functional_1984,UllrichBook} to calculate the electronic states of the system and its dynamics in the laser field.
The time-dependent Kohn-Sham potential 
\beq v_\mathrm{KS}[n](x, t) = v_\mathrm{ext}(x, t) + u[n](x, t) + v_\mathrm{xc}[n](x, t)\eeq 
is the sum of the external potential $v_\mathrm{ext}(x, t)$, 
the Hartree-potential $u[n](x, t)=\int n(x',t)/\sqrt{(x-x')^2+1}\,\diff x'$, where $n(x,t)$ is the electron density, 
and the exchange-correlation potential $v_\mathrm{xc}[n]$.
The topological effects studied in the present work are robust with respect to the choice of $v_\mathrm{xc}[n]$, therefore we use the simple adiabatic exchange-only local density approximation $v_\mathrm{xc}[n] \simeq - [3 n(x,t)/\pi]^{1/3}$.
The external potential $v_\mathrm{ext}(x, t) = v_\mathrm{ions}(x) - \mathrm{i} A(t) \frac{\partial}{\partial x}$ consists of the ionic potential $v_\mathrm{ions}(x)$ and the coupling to the laser field described by the vector potential $A(t)$ in ve\-loc\-i\-ty gauge and dipole approximation (with the $A^2$-term transformed away).

The implementation consists of two parts:
First, imaginary-time propagation is used to self-consistently determine  the occupied (and, if of interest, unoccupied) Kohn-Sham orbitals $\varphi_i$ without an external driver. %(e.g. laser).
In the second part, the influence of a laser pulse on the orbitals is simulated using real-time propagation.
The orbitals are propagated according to the time-dependent Kohn-Sham equation  
\beq \imagi\partial_t \varphi_i(x,t) = \left[-\frac{1}{2} \pabl{^2}{x^2} + v_\mathrm{KS}[n](x, t)  \right]\varphi_i(x,t) \eeq
with a split-operator Crank-Nicolson approximant~\cite{QSFQDBook}.
For the real-time propagation with time-dependent $u[n](x, t)$ and $v_\mathrm{xc}[n(x,t)]$, the Kohn-Sham equation is nonlinear, and a predictor-corrector scheme is used.
We restrict ourselves to spin-neutral systems in this work, so each Kohn-Sham orbital is occupied by a spin-up and a spin-down electron, and the total electron density reads $n(x,t)= 2 \sum_{i=1}^{N/2}|\varphi_i(x,t)|^2$.

\begin{figure}[htbp]
 \centering
 \includegraphics[width = 0.9 \columnwidth, height = \textheight, keepaspectratio]{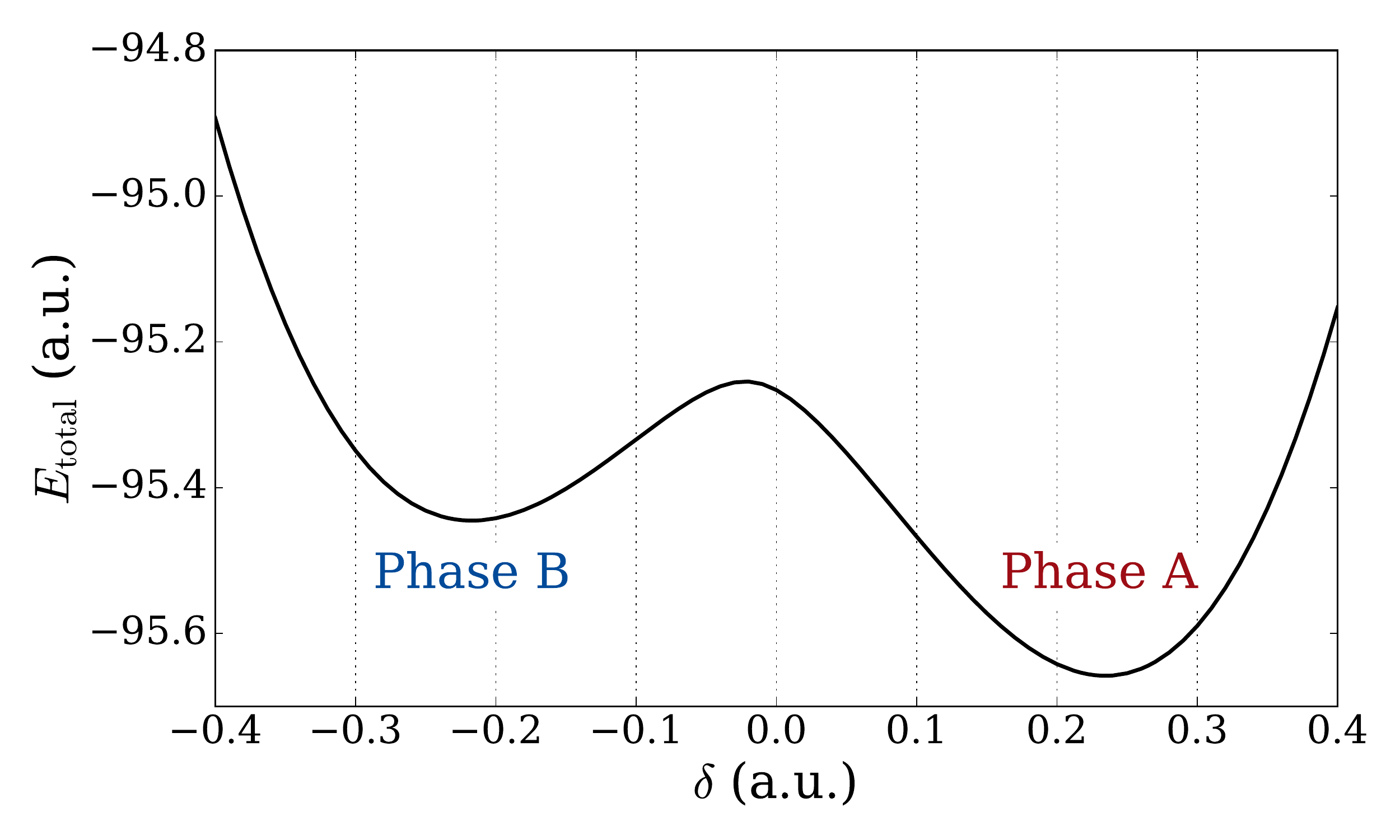}
 \caption{Total energy for $a=2$ and $N=100$ vs shift $\delta$. 
          \label{fig:total_energy}
         }
\end{figure}

Figure~\ref{fig:total_energy} shows the total energy for the chain with lattice constant $a=2$ (as used throughout this paper) and $N=100$ singly charged ions as a function of the shift $\delta$.
The total energy is calculated as
$
E_\mathrm{total} = E[\{ \varphi_{i}\}] + E_\mathrm{ii} 
$
with the static ion-ion energy $E_\mathrm{ii}  = \sum_{i=1}^N\sum_{j<i} [(x_j-x_i)^2 + 1]^{-1/2}$
 and the electronic energy  $E[\{ \varphi_{i}\}] = T_s[\{ \varphi_{i}\}] + E_\mathrm{ei}[n] + U[n] + E_\mathrm{xc}[n]$ where
$ T_s[\{ \varphi_{i}\} ] = -\sum_{i=1}^{N/2} \int\diff x\,  \varphi_{i}^*(x) \pabl{^2}{x^2} \varphi_{i}(x)$ is the kinetic energy, $ E_\mathrm{ei}[n] = \int v_\mathrm{ions}(x) n(x)\, \diff x$ is the electron-ion interaction energy, $U[n] = \halb \int u[n](x) n(x)\, \diff x$ is the Hartree energy, and
$ E_\mathrm{xc}[n] \simeq -\frac{3}{4} \left(\frac{3}{\pi}\right)^{1/3} \int n^{4/3}(x)\, \diff x$ is the exchange-correlation energy.
There is a global minimum in the phase~A ($\delta > 0$) regime at $\delta = 0.235$ and a local minimum in the phase~B ($\delta < 0$) at $\delta = -0.217$ \footnote{Figure~\ref{fig:total_energy} corrects panel (a) of Fig.~1 in~\cite{bauer_high-harmonic_2018} (which, however, does not affect any of the other results and the conclusions in~\cite{bauer_high-harmonic_2018}).}. The metallic case $\delta=0$ is energetically unfavorable (Peierls instability).

\begin{figure}[htbp]
 \centering
\includegraphics[width = 0.49 \columnwidth, height = \textheight, keepaspectratio]{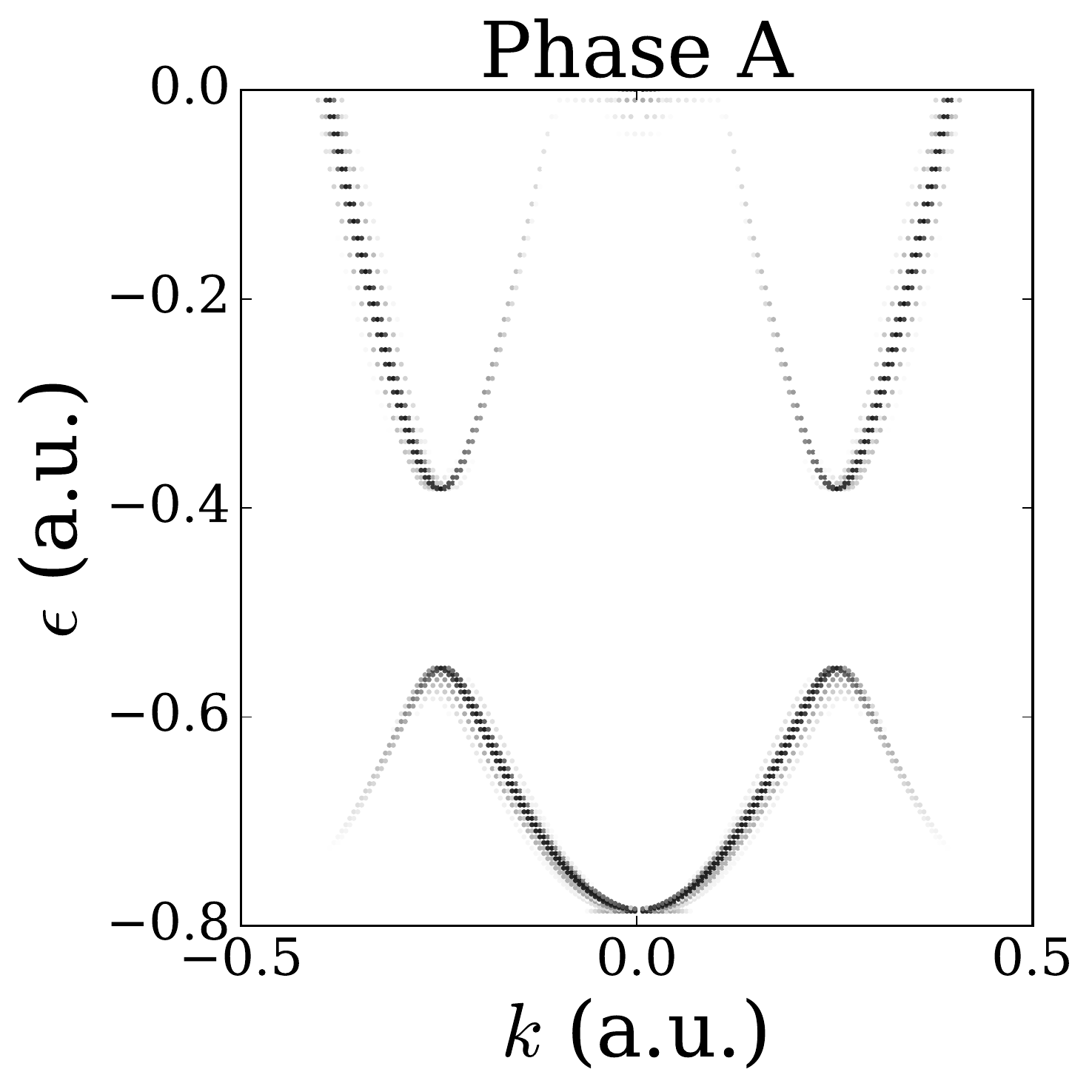}
\includegraphics[width = 0.49 \columnwidth, height = \textheight, keepaspectratio]{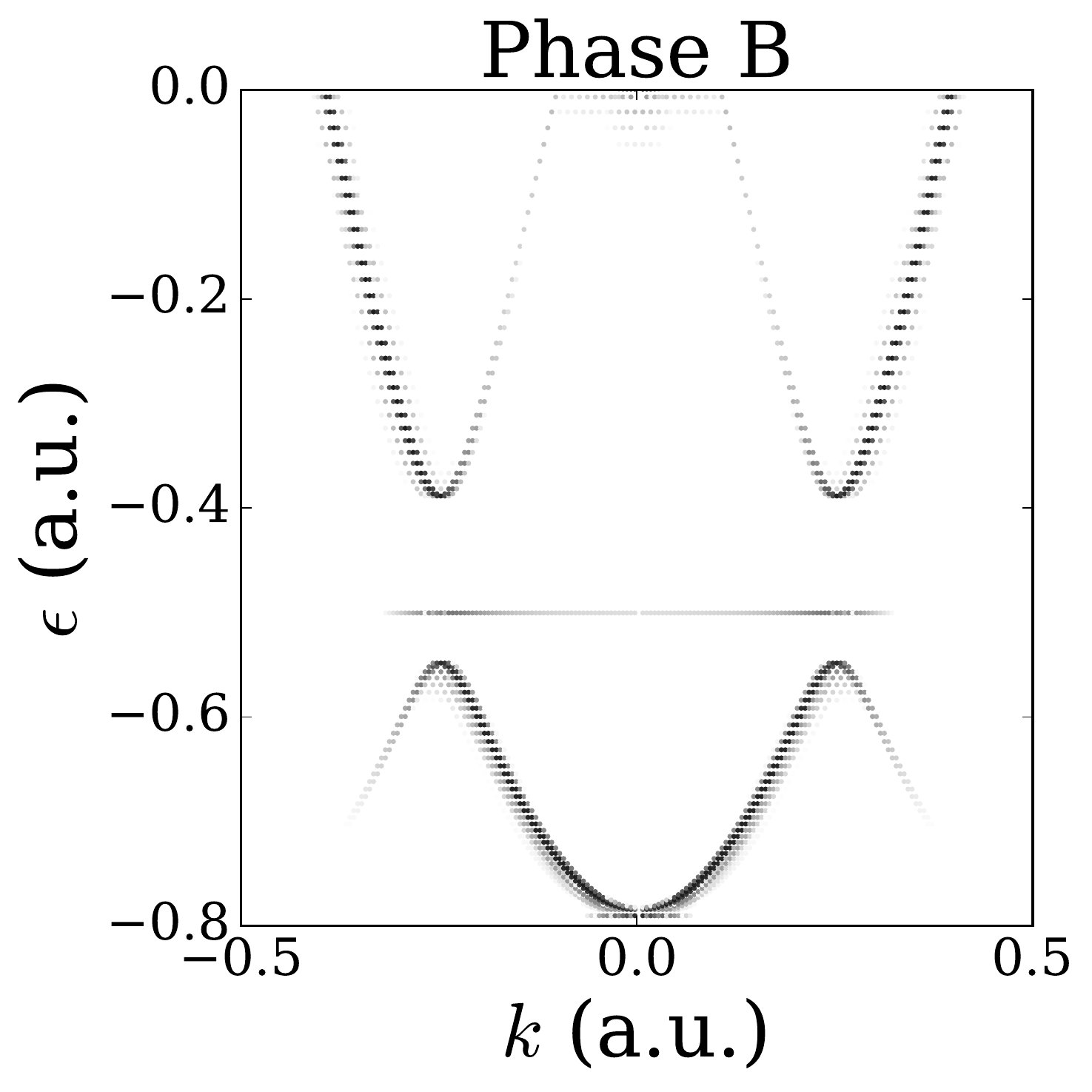}
 \caption{Band structures of phases A ($\delta = 0.235$) and B ($\delta = -0.217$). The lower band is fully populated. In phase B, one of the two degenerate edge states in the band gap is occupied.
          \label{fig:band_structures}
         }
\end{figure}

Figure~\ref{fig:band_structures} shows the band structures of the two phases at the local minima for $N=100$.
 We calculate band structures for finite chains by Fourier-transforming the position-space Kohn-Sham orbitals to $k$-space~\cite{PhysRevA.96.053418}.
Phase A is an insulator with a completely filled valence band and an unoccupied conduction band.
Two edge states are present in the band gap of phase~B.
They are degenerate and only one of them is occupied.
In position space, these states are localized on the edges of the chain, hence the name ``edge states.''

The laser pulse used in this work is 
an $N_\mathrm{cyc} = 5$ cycle sine-square pulse of frequency $\omega=0.0075$ (which corresponds to a wavelength of $\simeq \SI{6.1}{\um}$).
The vector potential 
reads
\beq A(t) = 
          A_0 \sin^2\left(\dfrac{\omega t}{2 N_\mathrm{cyc}}\right) \sin\omega t   \eeq 
for $ 0 < t < 2 \pi N_\mathrm{cyc}/\omega$ and zero otherwise.
The electronic dipole is proportional to
\beq X(t) = 2 \sum_{i=1}^{N / 2} \int x \big|\varphi_i(x, t)\big|^2 \,\mathrm{d} x\eeq 
and recorded during the laser pulse.
For some of the cases investigated in this work (e.g. during the transition from phase~B to phase~A, when the ions are not centered around zero), $X(0)\neq 0$ before the interaction with the laser.
To account for this, $X(0)$ is subtracted from $X(t)$ for all $t$, which only affects the harmonic component at frequency zero.
The topological features in the harmonic spectra are qualitatively independent of whether we use the dipole, velocity, current or acceleration to compute the spectra~\cite{bandrauk_quantum_2009,baggesen_dipole_2011}. 

In order to improve the dynamic range of the harmonic spectra, the dipoles are windowed by a 
Hann function~\cite{harris_use_1978}
$ w(t) %= \frac{1}{2} \Bigg(1 - \cos\left(\frac{2 \pi t}{N_t \Delta t - 1}\right)\Bigg)
       = \sin^2\left[{\pi t}/{(N_t \Delta t - 1)}\right]$, 
where $N_t$ is the number of outputs and $\Delta t$ the output interval.
The total dipole strength $D(\omega)$ is then calculated by taking the absolute square of the Fourier transform of the windowed dipole
\beq D(\omega) \propto \Big|\mathrm{FFT}\big[w(t) X(t)\big]\Big|^2.\eeq

\section{Robustness of topological effects on harmonic generation}\label{sec:robust}
Ref.~\cite{bauer_high-harmonic_2018} finds a difference of up to 14 orders of magnitude between harmonic spectra for the two topological phases.
The origin of this topological effect is attributed to the destructive interference of all the dipoles of the Kohn-Sham orbitals in phase~A due to the completely filled valence band.
In contrast, for phase~B, destructive interference is imperfect because of the only half-occupied edge states.
In the following, the remarkable robustness of that topological effect is illustrated.

\subsection{Size dependence} \label{sec:size}
\begin{figure}[htb]
 \includegraphics[width = 0.9 \columnwidth, height = \textheight, keepaspectratio]{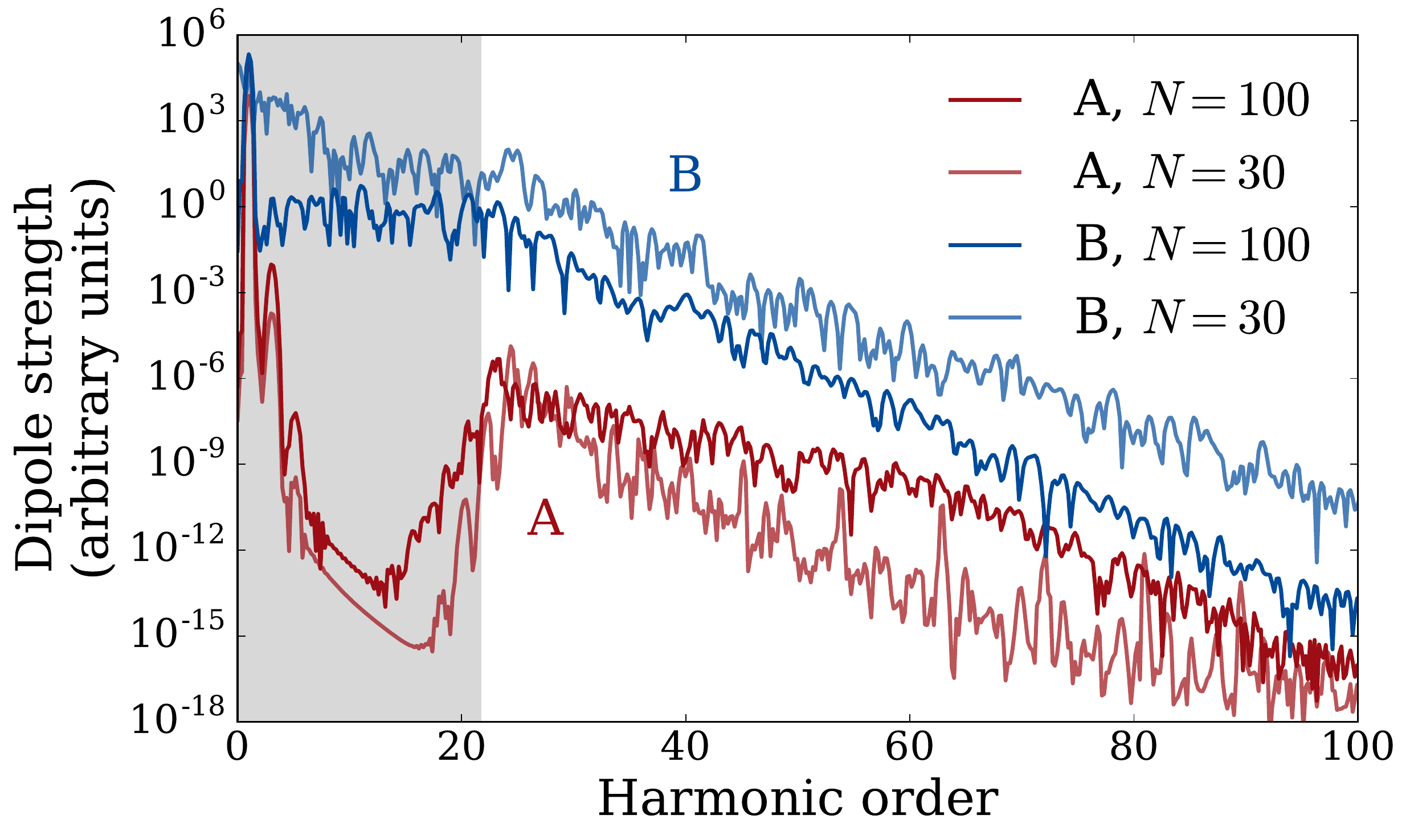}
 \caption{Harmonic spectra of phases A and B for $N = 100$ and $N = 30$ ions.
          The gray shade indicates the sub-band-gap regime of phase~A.
          \label{fig:HHG_fewer_ions}
         }
\end{figure}

The harmonic spectra for the two topological phases for $N=100$ ions and a vector potential amplitude $A_0=0.1$ (i.e., an intensity of $\simeq 2 \times 10^{10}\,$\Wcmcm) are shown in Fig.~\ref{fig:HHG_fewer_ions}.
The large difference in the harmonic yield observed already in~\cite{bauer_high-harmonic_2018} is clearly visible.

In order to investigate how many ions are needed to generate the characteristic spectra of the phases, the number of ions $N$ is decreased.
The effect of the chain length on the {\em high}-harmonic generation is investigated in Ref.~\cite{PhysRevA.97.043424}, although not with a focus on topological effects.
Here we focus on the sub-band-gap regime (i.e., harmonic-photon energies are smaller than the band gap of phase~A), where the large {\em qualitative} difference in Fig.~\ref{fig:HHG_fewer_ions} is observed.
 In the sub-band-gap regime, anharmonic intraband motion of the electrons is the origin of harmonics, and its dependence on the chain length is not obvious.
As the many-order-of-magnitude difference between phases A and B is an edge-state effect, one could expect that the ratio of surface (two ions in one dimension) to bulk (the other $N-2$ ions) is relevant, while  {\em topological} edge-state effects should be independent of $N$ as long as $N$ is large enough to yield an energy spectrum that resembles a band structure.
Indeed, increasing $N$ beyond 100 does not qualitatively change the harmonic spectra for both phases.

The harmonic spectra for $N=30$ are also shown in Fig.~\ref{fig:HHG_fewer_ions}.
Even though there are differences between the harmonic spectra for $N = 100$ and $N = 30$ ions in both phases, the large difference between the phases in the sub-band-gap regime remains qualitatively the same. 

Spectra for even smaller $N$ were examined as well.
The clear difference between the phases due to perfect and imperfect destructive interference deteriorates only when $N$ is so small ($N < 10$) that no band structure develops.
For the extreme case of $N=2$, the two phases correspond to two diatomic molecules with by $2\delta$ different internuclear distances where the notion of ``edge states'' does not make sense.

\subsection{Transition between phases}\label{sec:trans}
Topological equivalence is often popularly explained as the possibility to continuously deform an object without pinching holes in it or breaking it. Topological difference thus implies the impossibility of such a deformation without changing the topological invariant (e.g., the number of holes in an object).
In condensed matter physics, the situation is a bit more complex, as the deformations can be continuous in position space, but the topological invariant changes discontinuously in another, more abstract space.
Topological invariants are typically defined for bulk material as winding numbers (e.g., in the space spanned by the $2\times 2$ matrices of a tight-binding Bloch Hamiltonian as a function of lattice momentum $k$), integrals over the Berry curvature in $k$-space, or simply the number of levels below some properly defined zero energy~\cite{topinsRevModPhys.82.3045,topins,topinsshortcourse}. 

\begin{figure}[htbp]
%  \centering
 \includegraphics[width = 0.9 \columnwidth, height = 0.7\textheight, keepaspectratio]{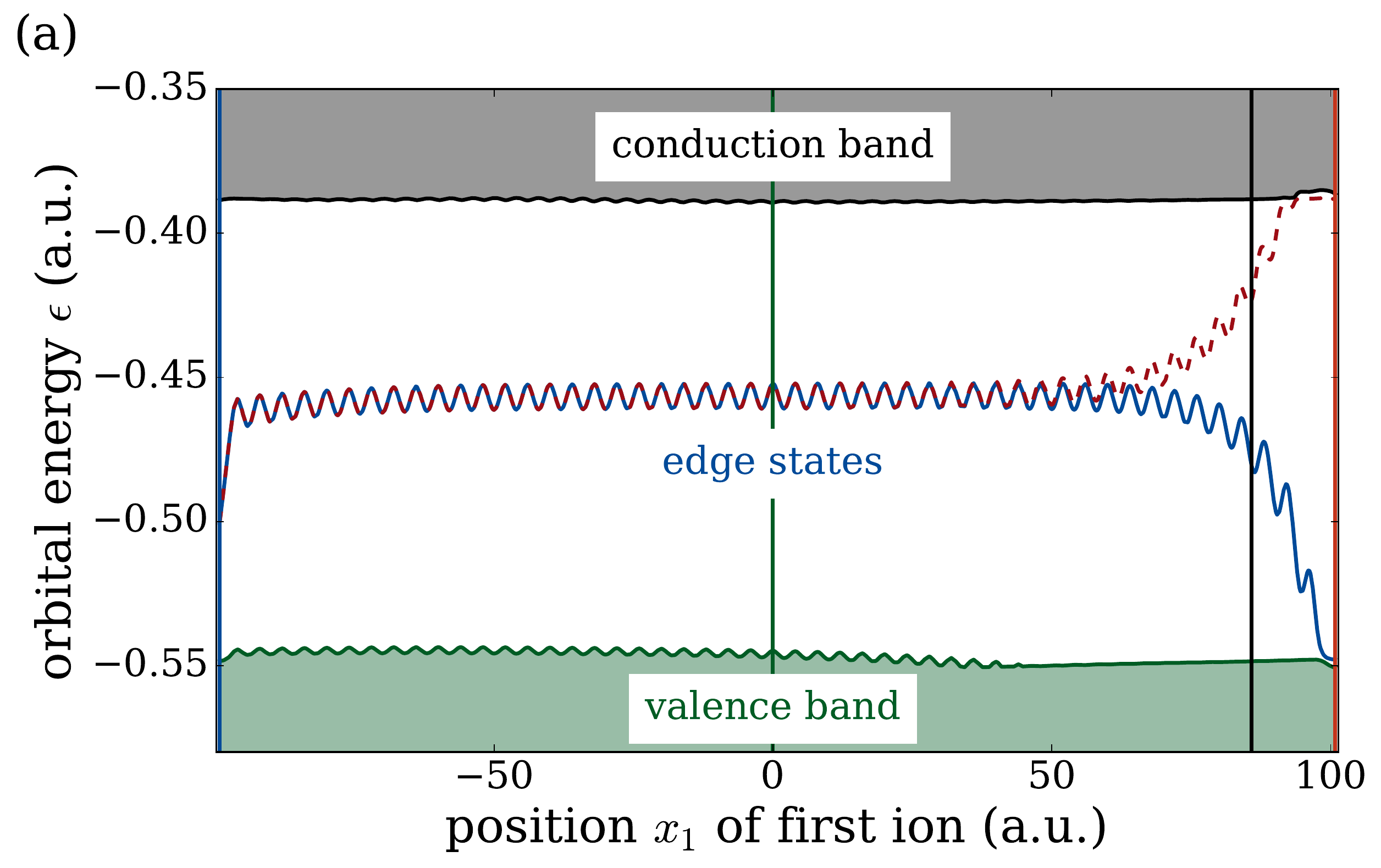}
 \includegraphics[width = 0.9 \columnwidth, height = 0.7\textheight, keepaspectratio]{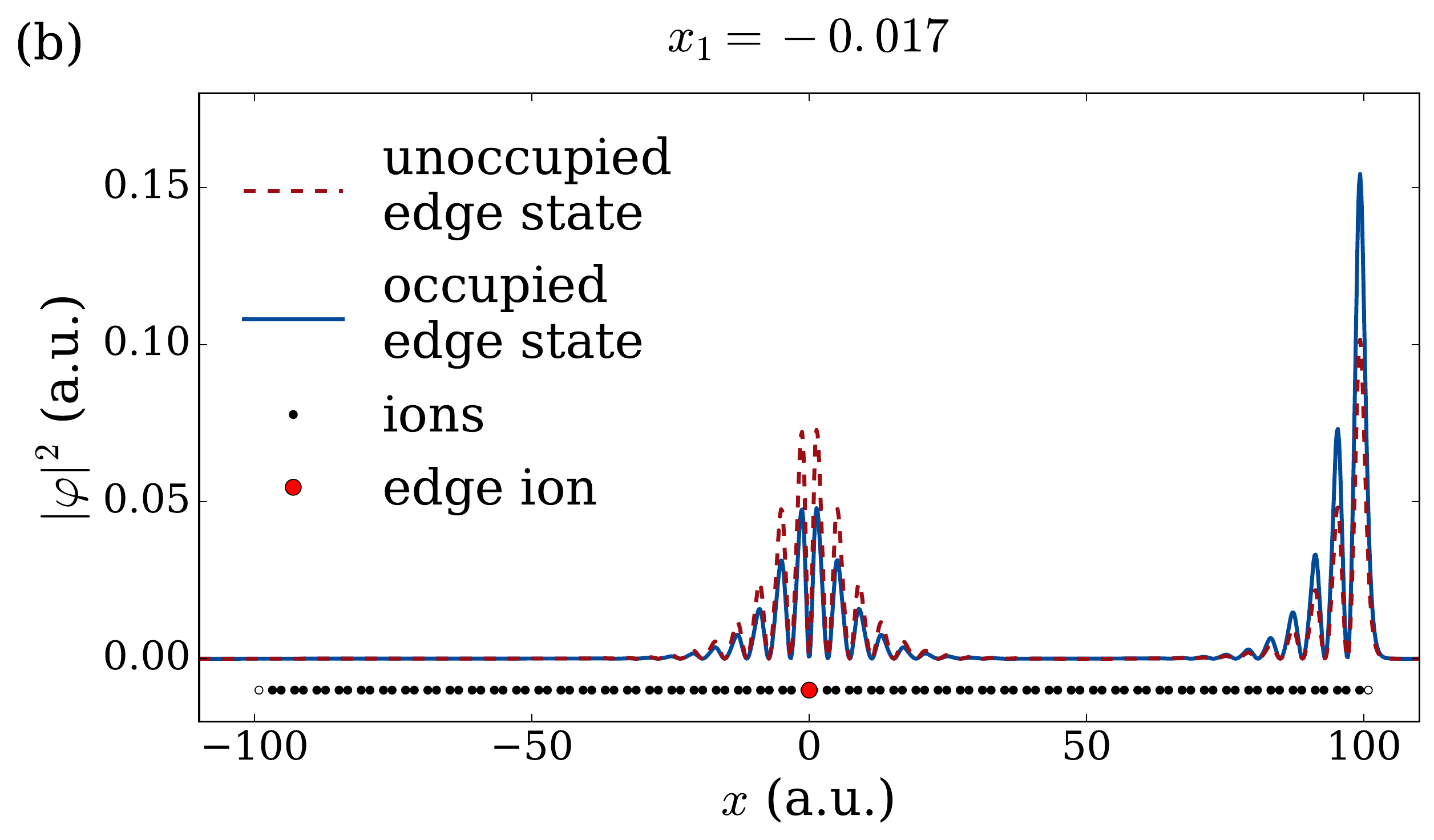}
 \includegraphics[width = 0.9 \columnwidth, height = 0.7\textheight, keepaspectratio]{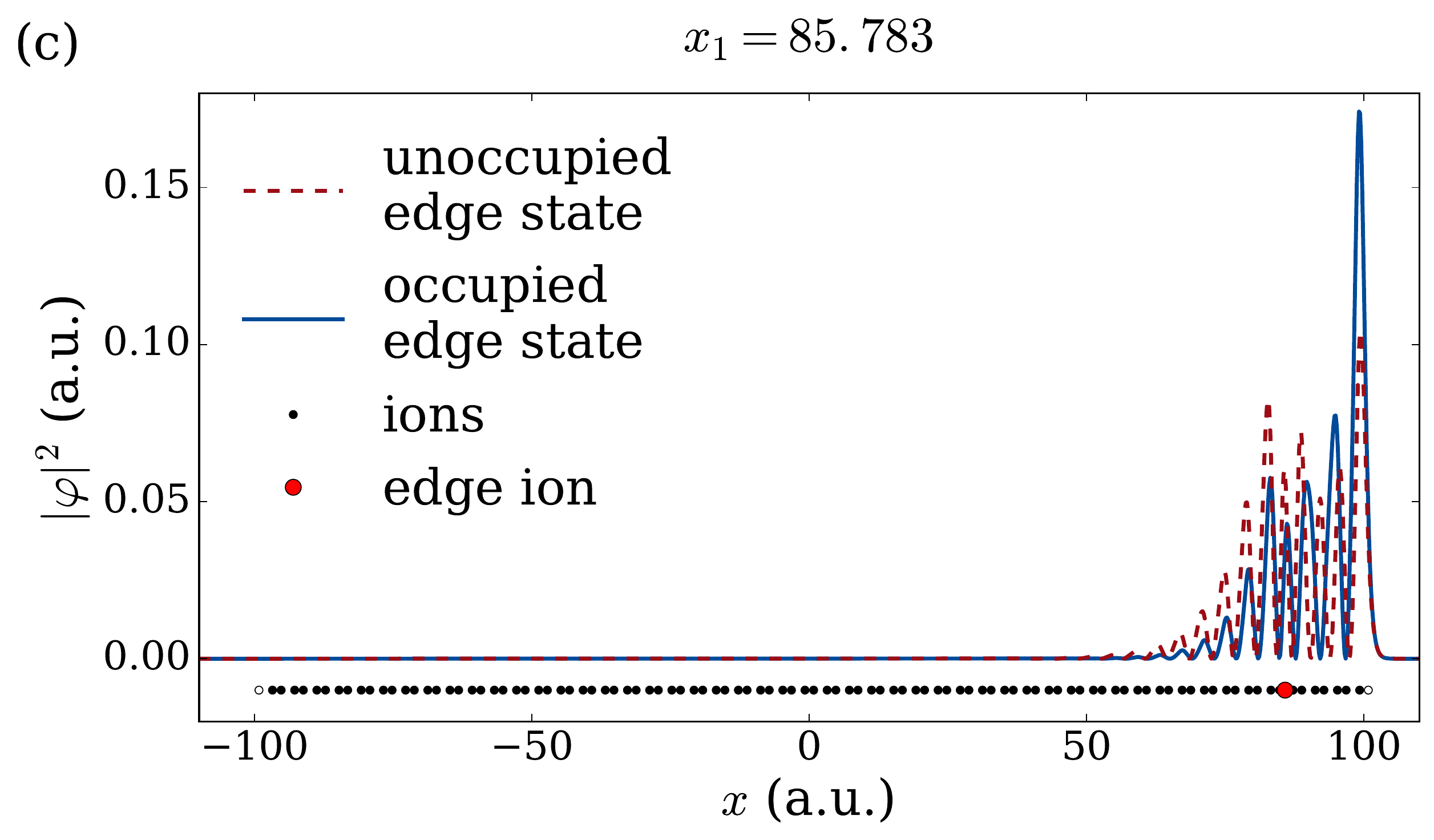}
 \caption{(a) Orbital energies during the transition from phase~B to phase~A (vertical lines indicate the positions of the spectra shown in Fig.~\ref{fig:HHG_moving_ion}), (b) edge state orbital densities for $x_1=-0.017$, and (c) $x_1=85.783$.
          }
 \label{fig:orbital_energies_moving_ion}
\end{figure}

We can shift continuously from phase~B to phase~A in position space by increasing the left edge ion's coordinate $x_1$ until it pairs up with the right edge ion to form a phase-A chain ($x_1 = x_{N+1} = x_{101}$). 
Figure~\ref{fig:orbital_energies_moving_ion} shows the orbital energies during the transition for $N=100$ as well as the edge state orbitals for two exemplary configurations. For each configuration, the self-consistent, stationary electron structure is calculated (i.e., we are not moving the edge ion in real time but use $x_1$ as an order parameter to describe a hypothetic, adiabatic phase transition).

One could expect that the degeneracy of the two edge-state energies is lifted as soon as the inversion symmetry is broken by an $x_1 \neq -x_{100}$. Instead, Fig.~\ref{fig:orbital_energies_moving_ion} shows that the edge-state levels remain degenerate until the edge-state orbitals start to overlap.
For example, for $x_1 \simeq 0$ (when the previously left-edge ion is moved to the center), the two contributions to the edge state orbitals from the right edge and the shifted left-edge ion are still well separated.
As a consequence, the energies of the two edge states are still degenerate, just slightly higher than in the original phase-B case.
At $x_1 \simeq 86$, the edge states are not degenerate anymore because the contributions on the shifted ion and the right edge overlap significantly.
The case $x_1 = 100.783$ almost corresponds to the perfect phase-A case. One of the two degenerate edge states turns into the highest state of the valence band, the other one into the lowest of the conduction band.

\begin{figure}[htbp]
%  \centering
 \includegraphics[width = 0.9 \columnwidth]{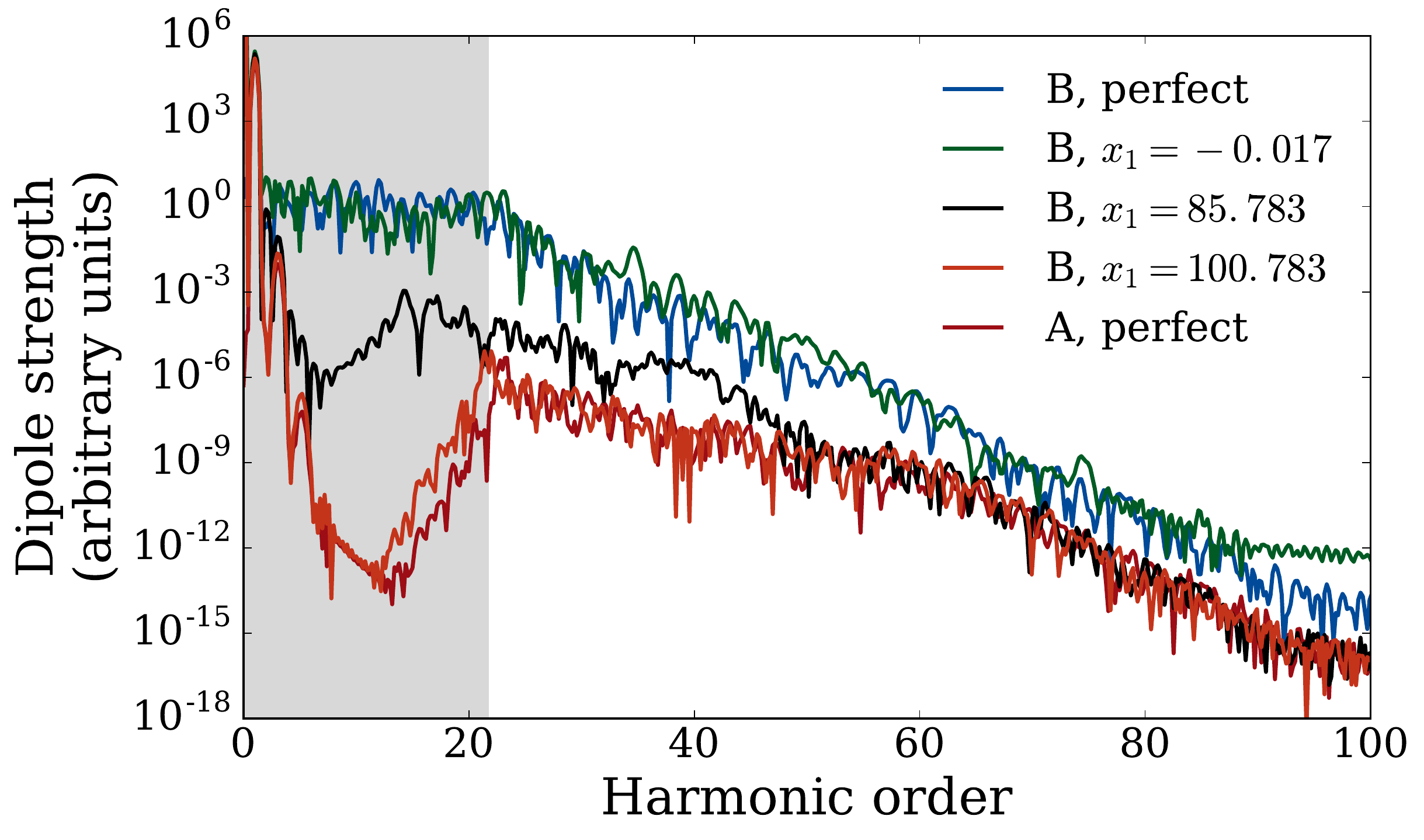}
 \caption{Harmonic spectra for phase~A, phase~B, and phase~B with $x_1 = -0.017$,  $x_1 = 85.7837$, and $x_1 = 100.783$ (these positions are indicated by vertical lines in Fig.~\ref{fig:orbital_energies_moving_ion}).
                    }
 \label{fig:HHG_moving_ion}
\end{figure}

Figure~\ref{fig:HHG_moving_ion} shows harmonic spectra for several $x_1$ during the transition, as indicated by vertical lines in Fig.~\ref{fig:orbital_energies_moving_ion}(a).
For $x_1 =-0.017$ the harmonic spectrum is very similar to the pure phase-B case because the band structure did not change qualitatively (the two edge states are still degenerate).
For  $x_1 = 85.783$, the spectrum already shows a phase-A-like dip, although a smaller one and at lower harmonic order because the energy difference between the two now separated edge states is smaller than the band gap in pure phase~A.
At $x_1 = 100.783$, the spectrum looks almost like the one for pure phase~A.
Small deviations are due to the fact that phase~B with $x_1 = 100.783$ corresponds to phase~A with $\delta = 0.217$ instead of $\delta = 0.235$ (where the total energy is minimal).

\subsection{Disorder in the ion positions} \label{sec:disorderpos}
Topological systems with disorder are a topic of current scientific interest (see, e.g.,~\cite{xu_stability_2006,Stutzer2018}).
The role of disorder in harmonic generation has been investigated as well~\cite{You2017}.
It has been found that disordered systems generate harmonics at least as well as ordered systems, showing that a band structure is not necessary for efficient harmonic emission from solids.

Disorder is introduced into our model system by randomly shifting the ion positions %according to
$x_i  \to x'_i = x_i' + r_i,$
mimicking disordered samples or finite temperature.
The random shifts $r_i$ follow a normal distribution with mean $0$ and standard deviation $\sigma_x$.

\begin{figure}[htb]
 \includegraphics[width = 0.3 \columnwidth, height = \textheight, keepaspectratio]{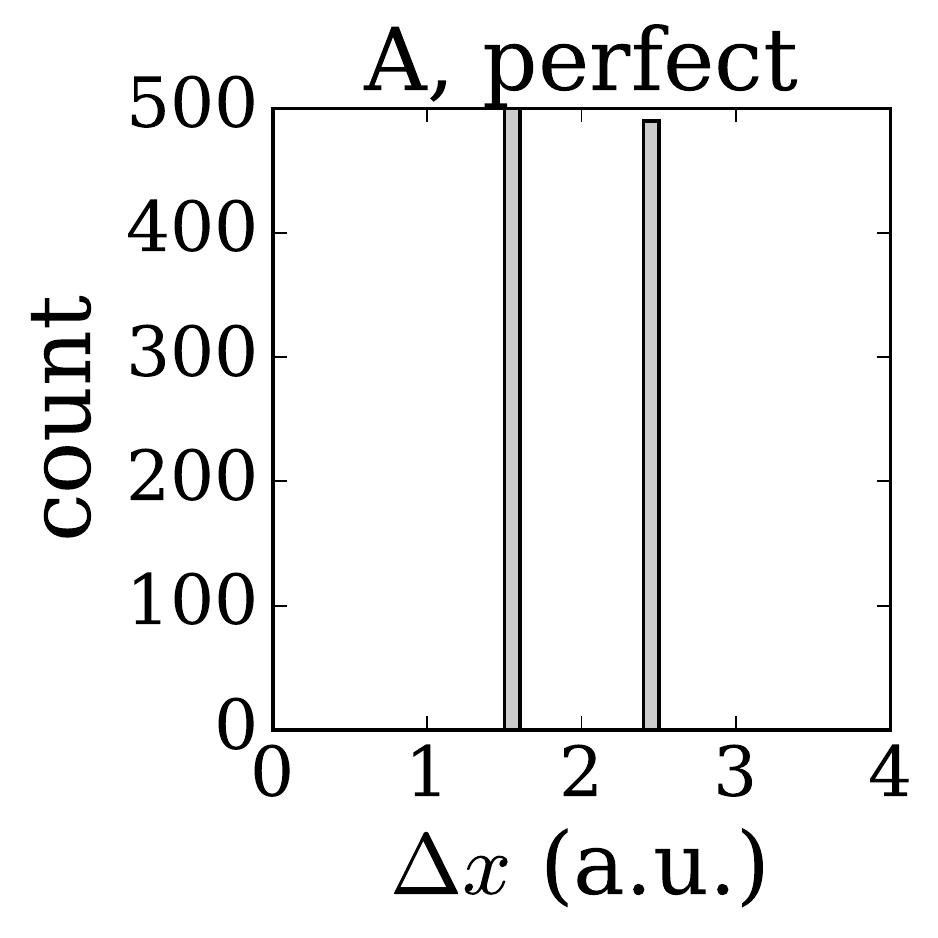}
 \includegraphics[width = 0.3 \columnwidth, height = \textheight, keepaspectratio]{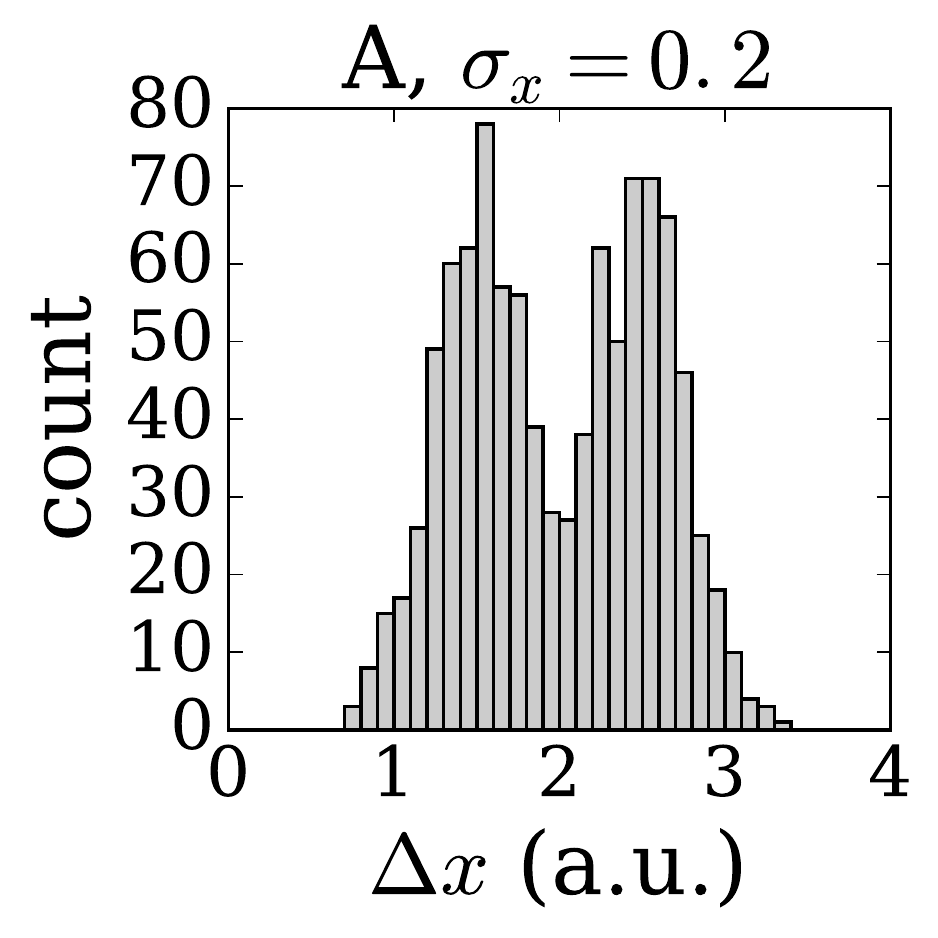}
 \includegraphics[width = 0.3 \columnwidth, height = \textheight, keepaspectratio]{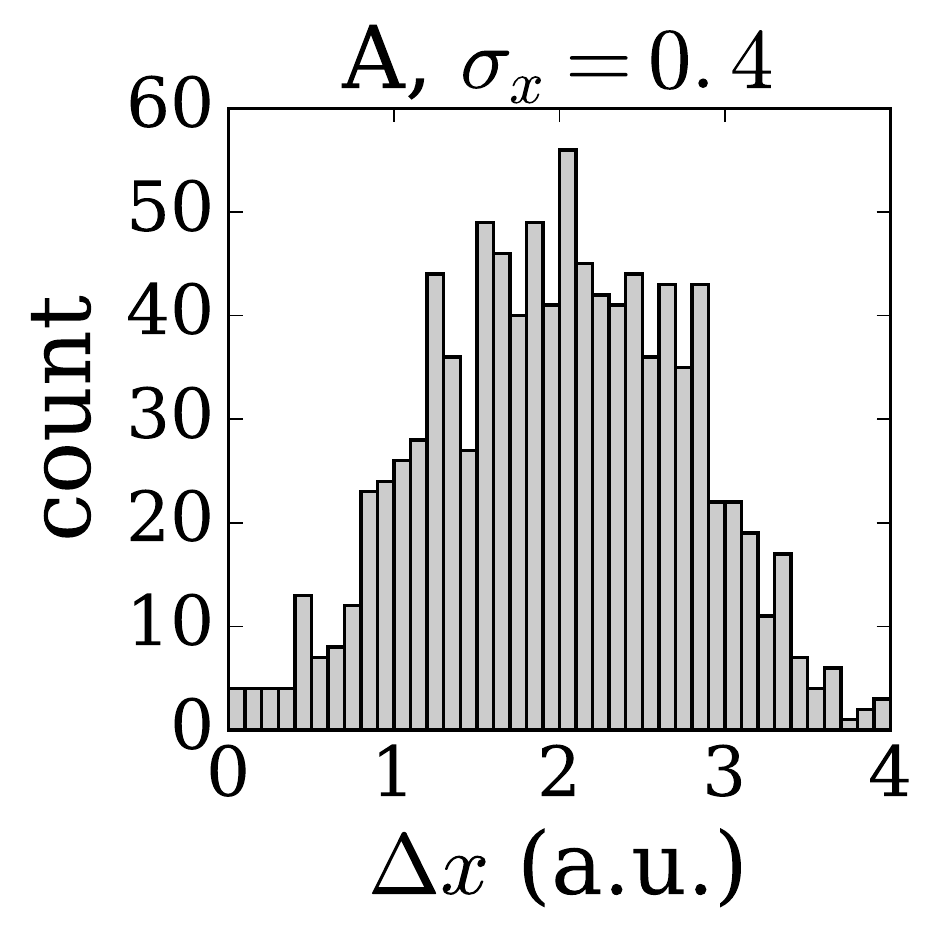}
 \includegraphics[width = 0.9 \columnwidth, height = \textheight, keepaspectratio]{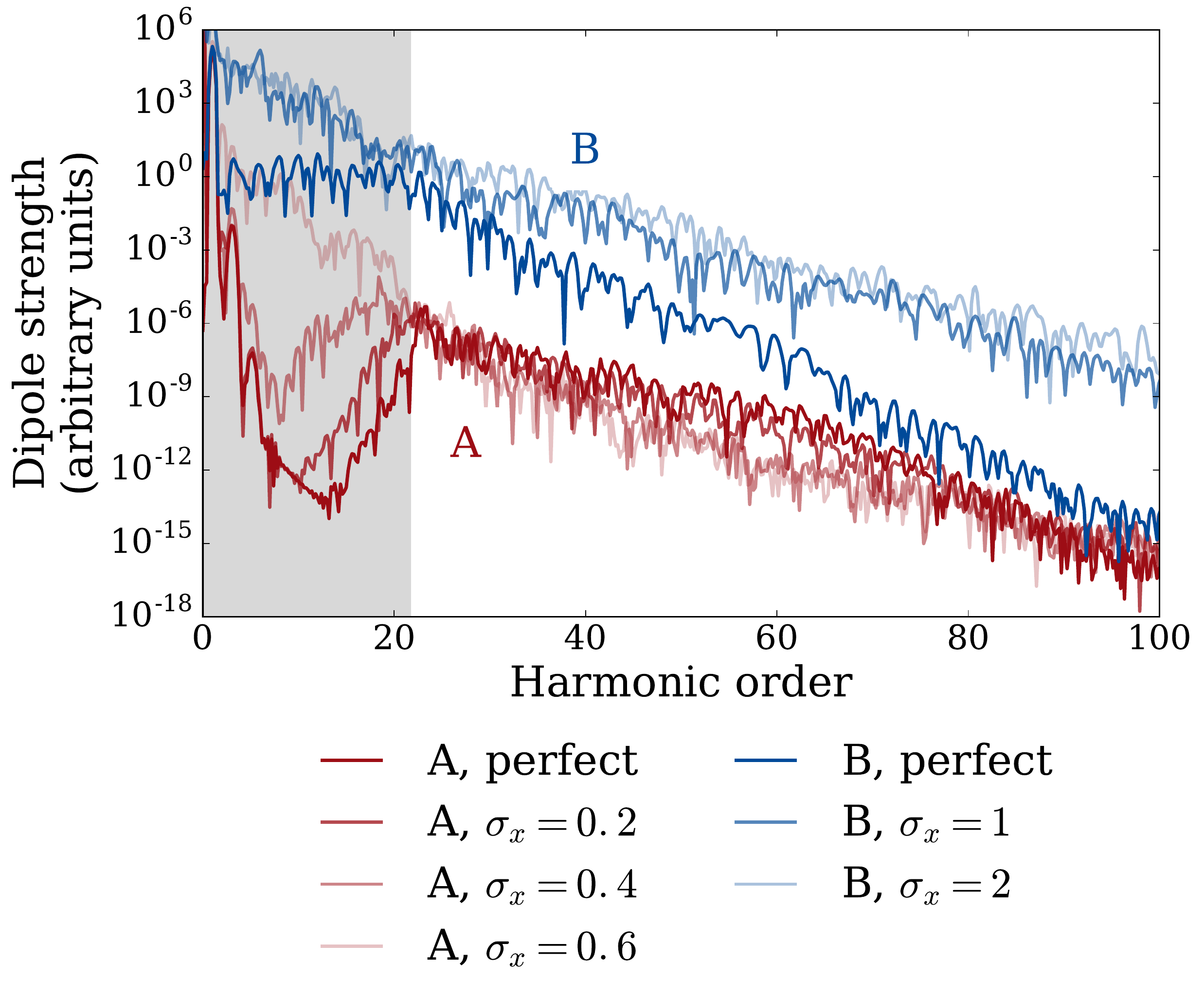}
 \caption{Top: Histograms of the internuclear distances $\Delta x$ for phase~A with random shifts.
          Bottom: Harmonic spectra for phases A and B for different standard deviations $\sigma_x$.
                   }
 \label{fig:high_harmonics_random_shifts}
\end{figure}

The harmonic spectra of phases A and B for different standard deviations $\sigma_x$ and histograms of the internuclear distances are shown in Figure \ref{fig:high_harmonics_random_shifts}.
In phase~A, the characteristic low harmonic yield below the band gap due to destructive interference of all the dipoles of Kohn-Sham electrons in the valence band disappears with increasing disorder.
This is because the bands are ``washed out'' by disorder, which effectively decreases the band gap.
However, the dip in the phase-A spectra survives up to surprisingly high $\sigma_x=0.4$, where the histogram of the internuclear distances does not show a two-peak distribution of a dimerized chain anymore.

The introduction of random shifts to the ion positions has little impact on the qualitative features of the harmonic spectrum from phase~B.
This is not surprising because there is no concerted destructive interference to be ruined by disorder in the first place. Therefore, we only show results for extreme disorder $\sigma_x = 1$ and $\sigma_x = 2$.
For such high values, many of the random shifts are already on the order of the lattice constant $a = 2$, i.e., ions are swapped. 
These spectra are thus examples for harmonic generation from purely random linear  chains.

\subsection{Disorder in the ion potentials} \label{sec:disorderpot}
Another way of introducing disorder is to randomly vary the smoothing parameters $\varepsilon_i$ in the Coulomb potentials $v_i(x) = - [(x-x_i)^2 + \varepsilon_i]^{-1/2}$ of the ions, which affects the depth of the potential. One may view this as  randomly changing the local ionization potentials of the atoms that constitute the chain. 
The $\varepsilon_i$ follow a normal distribution with mean $1$ and standard deviation $\sigma_\varepsilon$.
If a random $\varepsilon_i$ happens to be $<0$, it is set to $0$.

\begin{figure}[htb]
 \includegraphics[width = 0.85 \columnwidth]{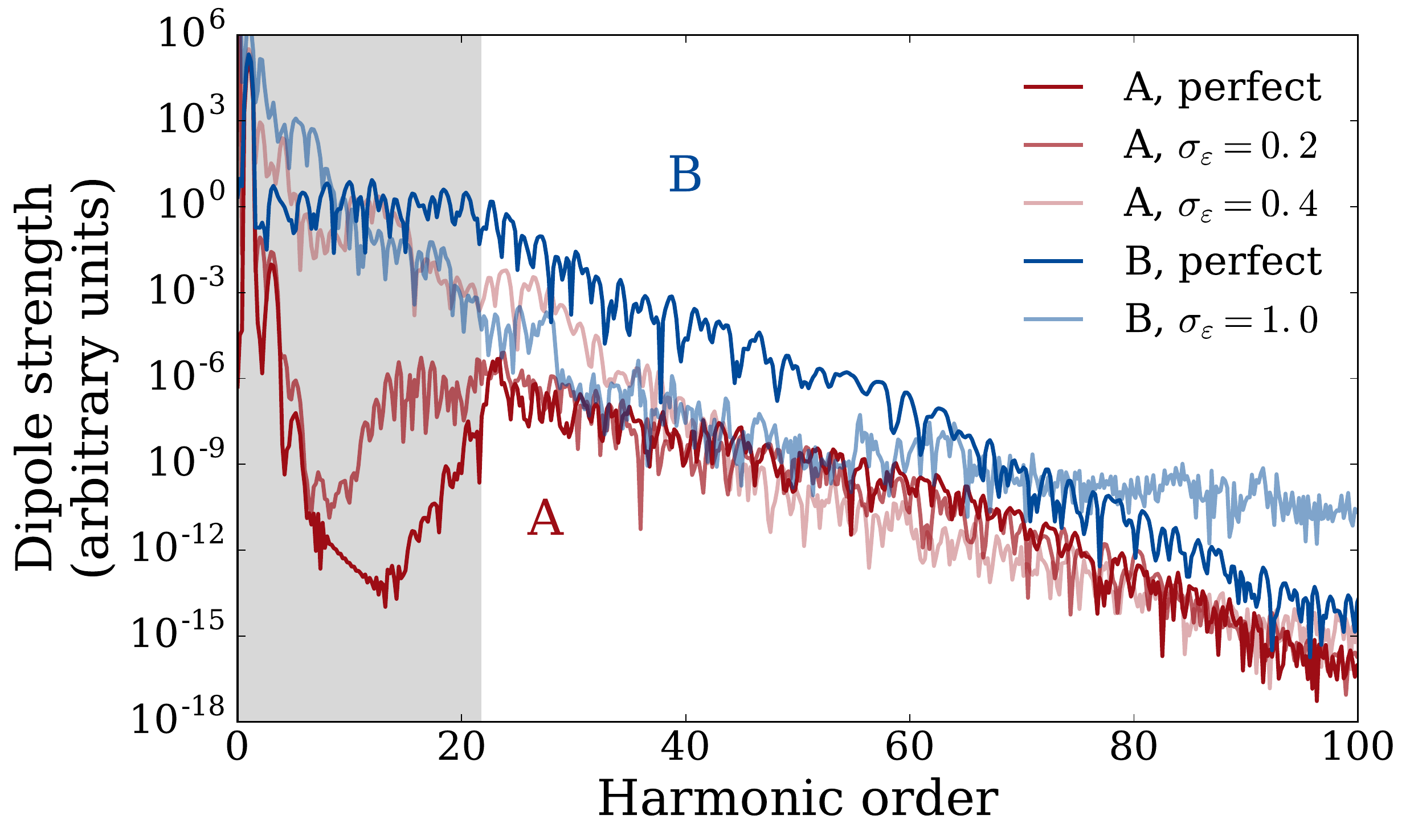}
 \caption{Harmonic spectra of phases A and B for different standard deviations $\sigma_\varepsilon$.
                    \label{fig:high_harmonics_random_coulomb}
         }
\end{figure}

The harmonic spectra of phases A and B for different standard deviations $\sigma_\varepsilon$ are shown in Figure \ref{fig:high_harmonics_random_coulomb}.
In phase~A, there are no large deviations in the above-band-gap regime (white background).
In the sub-band-gap regime, the characteristic low-harmonic yield disappears only for surprisingly large random variations (similar to the case of disordered ion positions in the previous subsection).

\section{Summary} \label{sec:summ}
Harmonic generation in different topological phases was investigated using a linear chain as a model system.
The harmonic yield was calculated with time-dependent density functional theory, taking the response of all electrons to the laser field into account.
The previously found low harmonic yield of phase~A (without topological edge states) compared to phase~B (with topological edge states) in the sub-band-gap regime was confirmed, and the robustness of this many-order-of-magnitude topological effect was studied with respect to (i) the size of the chain, (ii) a continuous transition between the two topological phases, and (iii) disorder.
When the number of ions $N$ in the system was lowered, both phases retained the characteristic features of their harmonic spectra down to very low $N\simeq 10$.
During the continuous transition from phase~B to phase~A by moving one of the edge ions to the other side, the degeneracy of the edge states and the harmonic spectrum remained of phase-B character until the shifted ion paired up with the opposite edge ion.
Random shifts of the ions and changes in the ionic potentials of the ions had to be surprisingly large in order to destroy the characteristic features of the harmonic spectra of both phases.

Regarding possible applications, robustness is probably the most important and striking asset of topological effects. Hence, it may be argued that the observed robustness in the present paper is not surprising.
However, it is important to point out that we observe this robustness for a time-dependent density-functional model of a strongly driven finite chain for which the topological invariant of the bulk Su-Schrieffer-Heeger chain does not exist.

\section*{Acknowledgments}
Inspiring discussions with Alexander Szameit are gratefully acknowledged.

\bibliography{bibliography}

\end{document}